\documentclass[lettersize,journal]{IEEEtran}
\usepackage{amsmath,amsfonts}
\usepackage{array}
\usepackage{caption} 
\usepackage{textcomp}
\usepackage{stfloats}
\usepackage{url}
\usepackage{verbatim}
\usepackage{graphicx}

\usepackage[T1]{fontenc}
\usepackage{cite}
\usepackage{subcaption}
\usepackage{epstopdf}
\usepackage{algorithm}
\usepackage{algpseudocode}

\usepackage{overpic}
\usepackage{amssymb}
\usepackage{amsthm}
\usepackage{array}
\usepackage{color}
\usepackage{url}
\usepackage{xspace}

\usepackage{xspace}

\theoremstyle{plain}

\newcommand{\vect}[1]{\mathbf{#1}}

\def\Htran{\mbox{\tiny $\mathrm{H}$}}
\def\Ttran{\mbox{\tiny $\mathrm{T}$}}
\def\imagunit{\mathsf{j}} 

\begin{document}

\title{Symbol-Level Precoding for Near-Field ISAC}


\author{\IEEEauthorblockN{\normalsize
    Nithin Babu, \IEEEmembership{~Member, ~IEEE},  Alva Kosasih, Christos Masouros,  ~\IEEEmembership{~Fellow, ~IEEE}, and  Emil Bj{\"o}rnson,~\IEEEmembership{~Fellow, ~IEEE}}
    
\thanks{This work was supported by project 6GMUSICAL part of the Smart Networks and Services Joint Undertaking (SNS JU) under the European Union’s Horizon Europe research and innovation programme under Grant Agreement No. 101139176 and the SUCCESS project funded by the Swedish Foundation for Strategic Research.}
\thanks{
    N. Babu and C. Masouros are with the Department of Electronic and Electrical Engineering, University College London, London, UK (E-mails: \{n.babu,c.masouros\}@ucl.ac.uk). A. Kosasih and E. Björnson are with the Division of Communication Systems, KTH Royal Institute of Technology, Stockholm, Sweden (E-mails: \{kosasih, emilbjo\}@kth.se). }
}

\maketitle

\begin{abstract} 
The forthcoming 6G and beyond wireless networks are anticipated to introduce new groundbreaking applications, such as Integrated Sensing and Communications (ISAC), potentially leveraging much wider bandwidths at higher frequencies and using significantly larger antenna arrays at base stations. This puts the system operation in the radiative near-field regime of the BS antenna array,  characterized by spherical rather than flat wavefronts. In this paper, we refer to such a system as near-field ISAC. Unlike the far-field regime, the near-field regime allows for precise focusing of transmission beams on specific areas, making it possible to simultaneously determine a target's direction and range from a single base station and resolve targets located in the same direction. This work designs the transmit symbol vector in near-field ISAC to maximize a weighted combination of sensing and communication performances subject to a total power constraint using symbol-level precoding (SLP). The formulated optimization problem is convex, and the solution is used to estimate the angle and range of the considered targets using the 2D MUSIC algorithm. The simulation results suggest that the SLP-based design outperforms the block-level-based counterpart. Moreover, the 2D MUSIC algorithm accurately estimates the targets' parameters.  
\end{abstract}


\vspace{-2mm}

\IEEEpeerreviewmaketitle

\section{Introduction}

The concept of combining sensing and communication functions into one system, known as  Integrated Sensing and Communications (ISAC), has gained significant interest from both academia and industry \cite{liu2022integrated}. Unlike traditional methods, where separate communication and sensing systems compete for the same radio resources, ISAC efficiently uses the same resources and hardware for both purposes.  
A key motivation for adopting ISAC 
includes a shift towards using higher frequencies and expanding the capabilities of the systems by incorporating more antennas.  

Utilizing a higher frequency and a larger Multiple-Input Multiple-Output (MIMO) dimension can potentially place users and targets in the radiative near-field regime of the base station (BS) antenna array. In this regime, the shape of the wavefront changes significantly. Rather than being flat, the wavefront becomes spherical due to its curvature being influenced by the size of the antenna array. As a rule-of-thumb, a wavefront can be considered flat (planar) when impinging on a small receiver only when the distance it travels is longer than the Fraunhofer distance, calculated as $2D^2/\lambda$, where $D$ is the array aperture and $\lambda$ is the wavelength \cite{2024_Kosasih_TWC}. If the wave travels shorter than the Fraunhofer distance, the spherical wavefront curvature is noticeable, indicating that the transmission occurs in the near-field regime.
 The Fraunhofer distance increases significantly with the carrier frequency ($f_c$) as well as the antenna array dimension. 

In the near-field regime, it is crucial to model both the direction and the distance from the source to the antenna array. This is necessary because the transmission beam is 
focused within specific spatial regions, a characteristic referred to as the beam focusing property \cite{2024_Kosasih_TWC}. 
The finite beamdepth enables simultaneous estimation of a target's range and azimuth angle in sensing applications. \color{black}According to \cite{liu2023integrated}, it is possible to localize a target in the far-field based on the appearance time, horizontal angle and azimuth angle information, albeit with additional processing cost. In contrast, this capability is inherently present in the near-field due to the spherical wavefront characteristic\color{black}. Furthermore, this capability allows the transmission of separate data to two distinct receivers aligned over the same angle from the transmitter, which is impossible  in the far-field operational regime. Despite its potential, this new research direction remains under-explored by prior literature.

The prior studies \cite{2023_Wang_Commlett, qu2023nearfield,elbir2023nearfield,luo2023beam,chen20236g} consider near-field ISAC. In \cite{2023_Wang_Commlett}, the authors consider block-level precoding (BLP) to minimize the Cram\'{e}r-Rao bound (CRB) for estimating the target distance and direction subject to a minimum communication rate requirement of each user. That study shows the performance gain of near-field ISAC over the far-field because of the available additional distance dimension. Similarly, \cite{qu2023nearfield} uses BLP to minimize a weighted combination of sensing and communication beamforming errors under a total power constraint. This enables a flexible trade-off between radar and communication objectives. Other existing studies, such as \cite{elbir2023nearfield,luo2023beam}, focus on designing precoders to mitigate beam-squint effects. An overview of the opportunities and challenges in near-field ISAC is given in \cite{chen20236g}.

The existing studies consider single-target scenarios and employ the BLP scheme to improve the considered performance metric. The BLP scheme aims to reduce interference and improve the network's signal-to-interference-plus-noise ratio (SINR). This is effective in instances characterized by a very large number of antennas or widely separated co-aligned users. In alternative scenarios, there is inevitable signal leakage between the users, restricting BLP's operational  SINR range. Unlike BLP, symbol level precoding (SLP) designs the transmission signal by jointly exploiting knowledge of the channel and the symbols to be transmitted \cite{9035662}. This enables the SLP to use co-channel interference constructively, aligning it with the desired signal at the receiver's end to improve SINR. Consequently, SLP facilitates operation in interference-limited scenarios, offering a promising alternative to BLP. 

This paper investigates the potential of SLP to enhance both estimation and communication performance in a near-field ISAC system with multiple targets and users. We develop an optimization framework to design the transmit symbol vector, aiming to maximize a weighted combination of sensing and communication metrics. Sensing performance is evaluated using the derived \color{black}Cram\'{e}r-Rao bound (CRB)\color{black}, while maximization of the minimum SINR is the communication objective. The optimization problem is convex, and the resulting solution is employed by a 2D MUSIC algorithm to estimate the angles and ranges of the targets. 
The primary contribution is the application of SLP to design the transmit symbol vector of a near-field ISAC, considering multiple targets and users. This differs from prior studies, such as \cite{2023_Wang_Commlett} and \cite{qu2023nearfield}, which predominantly utilize BLP in single-target, multi-user scenarios. Simulation results confirm the superior performance of SLP compared to BLP.

\section{System Model}

\begin{figure}
    \centering
    \includegraphics[width=0.7\columnwidth]{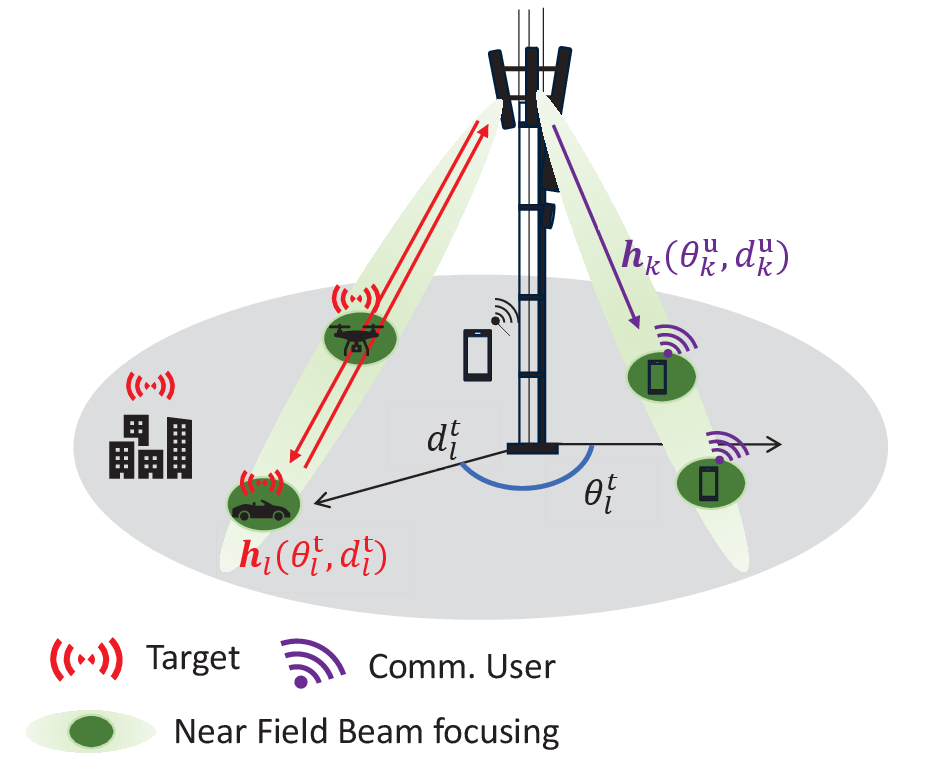}
    \caption{Illustration of the considered near-field ISAC system.}
    \label{fig:system_model}
    \vspace{-5mm}
\end{figure}

We consider an ISAC system wherein a BS is equipped with an $N$-element transmit Uniform Linear Array (ULA) and an $N$-element receive ULA, respectively, as depicted in Fig. \ref{fig:system_model}. The BS provide data communication to $K$ single-antenna user equipments (UEs)
while detecting $L$ targets, \color{black}all located in the near-field regime of the BS antenna array\color{black}.   
\vspace{-3mm}
\subsection{Near-Field Channel Model}

Let the BS's transmitting and receiving antennas be geometrically arranged along the $x$-axis with a half-wavelength spacing and the reference element be located at $(0,0,0)$. The $n^{\text{th}}$ antenna element is centered at $(\bar{x}_n,0,0)$ with
\begin{equation}
\bar{x}_n = \underbrace{\left(n-\frac{N+1}{2} \right)}_{\delta_n} \underbrace{\frac{\lambda}{2}}_{\Delta},
\end{equation}
 where $\delta_n$ and $\Delta$ are the index of $n$-th antenna element and spacing between two antenna elements, respectively.




The channel between the $n^{\text{th}}$ antenna element and the $k^{\text{th}}$ user located at a distance of $d_k$ at an angle of $\theta_k$ with respect to the reference antenna element is represented as 
\begin{equation}\label{eq:ch_resp}
  h_{n}^k(d_k,\theta_k) = {\sqrt{\beta_{n,k}}}   e^{-\imagunit \frac{2\pi}{\lambda}r_{n}^k},
\end{equation}
where $r^{k}_{n}=\sqrt{d^{2}_{k}+(\delta_{n}\Delta)^{2}-2d_{k}\Delta\delta_{n}\cos(\theta_{k})}$ \cite{qu2023nearfield}.  If the Euclidean distance between a user/target and the BS is less than the Fraunhofer array distance of the BS antenna array,  then the user/target is considered to be in the radiative near-field region (Fresnel region) of the BS array. The free-space path loss between antenna $n$ and user $k$ ($\beta_{n,k}$), which can be approximated as \color{black}\cite{2024_Kosasih_Arxiv}\color{black}
\begin{equation}\label{eq:beta_approx}
   \beta_{n,k}  \approx   \beta_{k} = 
   { \frac{\lambda^2}{16 \pi}}\frac{\sin(\theta_k)}{d_k^{2}}.
\end{equation}
The approximation holds when the propagation distance is larger than twice the array aperture \cite{2020_Björnson_JCommSoc} so that the spherical amplitude variations over the wavefront are negligible but not the phase variations.
 Using \eqref{eq:ch_resp} and \eqref{eq:beta_approx}, we  can model the near-field channel vector to the user $k$ as
\begin{align}\label{ori_arr_resp}
    \vect{h}_k (d_k,\theta_k)  &= {\sqrt{\beta_{k}}}  \left[ e^{-\imagunit \frac{2\pi}{\lambda}r_{1}^k}, \dots, e^{-\imagunit \frac{2\pi}{\lambda}r_{N}^k} \right]^{\Ttran}.
\end{align}
\subsection{Sensing Model}
\color{black}The BS transmits a narrowband symbol matrix, $\mathbf{X}=\left[\mathbf{x}_{1},\mathbf{x}_{2},...,\mathbf{x}_{S}\right] \, \in \mathbb{C}^{N\times S}$, with $S > N$ being the length of the radar pulse communication frame\color{black}. The transmitted symbols get reflected by the targets, and the resulting echo symbol matrix received by the BS from the targets is given as
\begin{IEEEeqnarray}{rCl}
\mathbf{Y}^{\mathrm{R}} &=& \sum_{l=1}^{L}b_{l}\mathbf{a}(d_{l},\theta_{l})\mathbf{v}^{\Ttran}(d_{l},\theta_{l})\mathbf{X}+ \mathbf{Z}^{\mathrm{R}},\label{yr} 
\end{IEEEeqnarray}
where $\mathbf{Y}^{\mathrm{R}}   = [\mathbf{y}^{\mathrm{R}}_1, \mathbf{y}^{\mathrm{R}}_2, \dots, \mathbf{y}^{\mathrm{R}}_S] \in \mathbb{C}^{N\times S} $, $\mathbf{a}(d_{l},\theta_{l})$ and $\mathbf{v}(d_{l},\theta_{l})$ are the receive and transmit near-field array response vectors in \eqref{ori_arr_resp},   $(\cdot)^{\mathrm{T}}$ denotes the transpose operation, and $\mathbf{Z}^{\mathrm{R}}\in \mathbb{C}^{N \times S}$ is an additive white complex Gaussian noise (AWGN) matrix where the variance of each entry is $\sigma^{2}_{\mathrm{R}}$. Additionally, $b_l=b^{\mathcal{R}}_{l}+ j b^{\mathcal{I}}_{l}$ represents the $l^{\text{th}}$ target's complex amplitude and its magnitude is the radar cross section (RCS).

\vspace{-3mm}
\subsection{Sensing Performance Metric}
Let the vector of parameters to be estimated be $\boldsymbol{\zeta}=\{\boldsymbol{\theta}^{\mathrm{t}},\mathbf{d}^{\mathrm{t}},\mathbf{b}^{\mathcal{R}}_{l},\mathbf{b}^{\mathcal{I}}_{l}\}$. Here, $\boldsymbol{\theta}^{\mathrm{t}}=\{{\theta}_{l}^{\mathrm{t}}\}_{l=1}^{L}$, $\boldsymbol{d}^{\mathrm{t}}=\{{d}_{l}^{\mathrm{t}}\}_{l=1}^{L}$, $\mathbf{b}^{\mathcal{R}}=\{{b}^{\mathcal{R}}_{l}\}_{l=1}^{L}$, and $\mathbf{b}^{\mathcal{I}}=\{{b}^{\mathcal{I}}_{l}\}_{l=1}^{L}$. \color{black}The error variance is lower-bounded by the CRB, which is given by the inverse of the Fisher Information Matrix (FIM) \cite{dong2022sensing}\color{black}. The received signal in \eqref{yr} is a multivariate Gaussian random variable with mean $\boldsymbol{\mu}=\sum_{l=1}^{L}b_{l}\mathbf{a}(d_{l},\theta_{l})\mathbf{v}^{\Ttran}(d_{l},\theta_{l})\mathbf{X}$ and covariance matrix $\mathbf{C}=\sigma^{2}_{\mathrm{R}}\mathbf{I}_{N}$ \cite{babu2023multi}. Since we need to estimate a total of $|\boldsymbol{\zeta}|$ parameters, the FIM is a $|\boldsymbol{\zeta}|\times|\boldsymbol{\zeta}|$ matrix with the entries 
\begin{align}
{\left[\mathbf{F}\right]}_{\zeta_{l},\zeta_{p}} = 2\cdot \text{Re}\left\lbrace\mathrm{tr}\left( \frac{\partial \boldsymbol{\mu}^{{\Htran}}}{\partial {\zeta}_{l}} \mathbf{C}^{-1} \frac{\partial \boldsymbol{\mu}}{\partial {\zeta}_{p}} \right)\right\rbrace,\quad \forall \zeta_{l} \in \zeta,
\end{align}
where $\frac{d\boldsymbol{\mu}}{d {\zeta}_{l}}$ is the partial derivative of $\boldsymbol{\mu}$ with respect to the parameter $\zeta_{l} \in \boldsymbol{\zeta}$. Here,
\begin{align}
   & \frac{\partial \boldsymbol{\mu}}{\partial {\theta}^{\mathrm{t}}_{l}} = \left(\dot{\mathbf{a}}_{l,\theta^{\mathrm{t}}}{\mathbf{v}}^{\Ttran}_{l}+{\mathbf{a}}_{l} \dot{\mathbf{v}}^{\Ttran}_{l,,\theta^{\mathrm{t}}}\right)\mathbf{X}, \label{tmucbf}\\
   & \frac{\partial \boldsymbol{\mu}}{\partial {d}^{\mathrm{t}}_{l}} = \left(\dot{\mathbf{a}}_{l,d^{\mathrm{t}}}{\mathbf{v}}^{\Ttran}_{l}+{\mathbf{a}}_{l} \dot{\mathbf{v}}^{\Ttran}_{l,d^{\mathrm{t}}}\right)\mathbf{X}, \\
   & \frac{\partial \boldsymbol{\mu}}{\partial {b}^{\mathcal{R}}_{l}} = {\mathbf{a}}_{l}{\mathbf{v}}^{\Ttran}_{l}\mathbf{X},  \frac{d \boldsymbol{\mu}}{d {b}^{\mathcal{I}}_{l}} = \imagunit{\mathbf{a}}_{l}{\mathbf{v}}^{\Ttran}_{l}\mathbf{X}, \label{dmucbf}   
\end{align} 
where $\mathbf{a}_{l}$ and $\mathbf{v}_{l}$ stand for $\mathbf{a}(\theta^{\mathrm{t}}_{l},d^{\mathrm{t}}_{l})$ and $\mathbf{v}(\theta^{\mathrm{t}}_{l},d^{\mathrm{t}}_{l})$, respectively. Here, $\dot{\mathbf{d}}_{l,\zeta}$ is the derivative of $\mathbf{d}_{l}$ with respect to $\zeta_{l}$ for $\mathbf{d}_{l} \in \{\mathbf{a}_{l},\mathbf{v}_{l} \}$. Using \eqref{first}-\eqref{last}, the FIM can be represented as
\begin{align}
   \mathbf{F}=2\begin{bmatrix}\text{Re}\left[\mathbf{F}\right]_{\boldsymbol{\theta}^{\mathrm{t}}\boldsymbol{\theta}^{\mathrm{t}}}&\text{Re}\left[\mathbf{F}\right]_{\boldsymbol{\theta}^{\mathrm{t}}\mathbf{d}^{\mathrm{t}}}&\text{Re}\left[\mathbf{F}\right]_{\boldsymbol{\theta}^{\mathrm{t}}\mathbf{b}}&-\text{Im}\left[\mathbf{F}\right]_{\boldsymbol{\theta}^{\mathrm{t}}\mathbf{b}}\\
   \text{Re}\left[\mathbf{F}\right]^{\mathrm{T}}_{\boldsymbol{\theta}^{\mathrm{t}}\mathbf{d}^{\mathrm{t}}}&\text{Re}\left[\mathbf{F}\right]_{\mathbf{d}^{\mathrm{t}}\mathbf{d}^{\mathrm{t}}}&\text{Re}\left[\mathbf{F}\right]_{\mathbf{d}^{\mathrm{t}}\mathbf{b}}&-\text{Im}\left[\mathbf{F}\right]_{\mathbf{d}^{\mathrm{t}}\mathbf{b}}\\
   \text{Re}\left[\mathbf{F}\right]^{\mathrm{T}}_{\boldsymbol{\theta}^{\mathrm{t}}\mathbf{b}}&\text{Re}\left[\mathbf{F}\right]^{\mathrm{T}}_{\mathbf{d}^{\mathrm{t}}\mathbf{b}} & \text{Re}\left[\mathbf{F}\right]_{\mathbf{b}\mathbf{b}}&-\text{Im}\left[\mathbf{F}\right]_{\mathbf{b}\mathbf{b}}\\
   -\text{Im}\left[\mathbf{F}\right]^{\mathrm{T}}_{\boldsymbol{\theta}^{\mathrm{t}}\mathbf{b}}&-\text{Im}\left[\mathbf{F}\right]^{\mathrm{T}}_{\mathbf{d}^{\mathrm{t}}\mathbf{b}}&-\text{Im}\left[\mathbf{F}\right]^{\mathrm{T}}_{\mathbf{b}\mathbf{b}}&\text{Re}\left[\mathbf{F}\right]_{\mathbf{b}\mathbf{b}}\\
   \end{bmatrix}. \label{fmatrix}
\end{align}
\vspace{-8mm}
\subsection{Communication Model}
 The received  signal $\mathbf{y}_k \in \mathbb{C}^{ S }$ can be expressed as
\begin{equation}
    \mathbf{y}_k = \vect{h}^{\Ttran}_{k}\mathbf{X}+ \vect{n}_k,
\end{equation}
where $\vect{h}^{\Ttran}_{k}=\vect{h}^{\Ttran}_{k}(d_k,\theta_k)$ and $\vect{n}_k  \in \mathbb{C}^{S}$ is the AWGN noise vector with the variance of each entry being $\sigma^{2}_{\rm C}$. Here, we consider maximizing the minimum communication SINR as the communication performance metric. In contrast to \cite{chen20236g,2023_Wang_Commlett}, which consider BLP to maximize communication performance, we consider SLP. 
\subsubsection{SINR constraint using SLP}
\begin{figure}
    \centering
    \includegraphics[width=0.49\columnwidth]{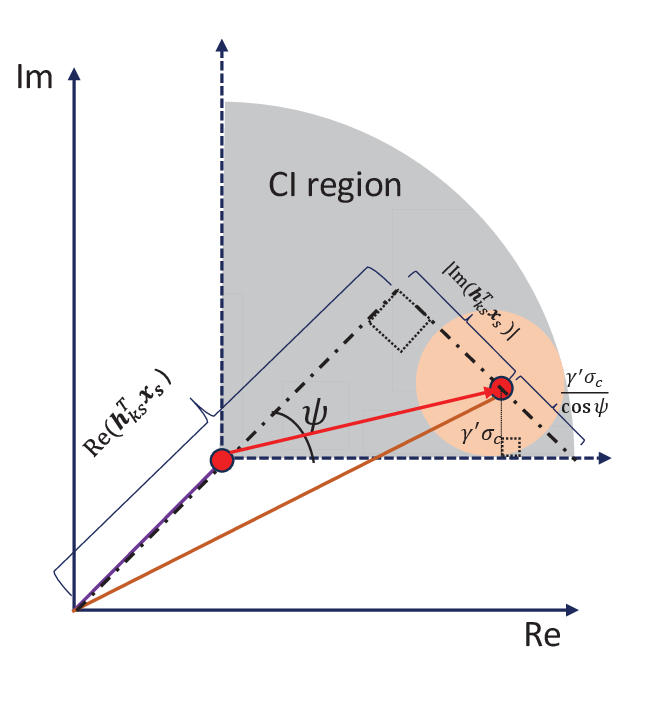} \vspace{-4mm}
    \caption{Geometric representation of SINR constraint.} \vspace{-3mm}
    \label{fig:slp}
    \vspace{-3mm}
\end{figure}
In this study, we use the SLP method to design the transmission symbol vector, effectively improving both sensing and communication metrics. The SLP enhances the SINR by aligning the interfering signals with the intended signal at the reception point. As shown in Fig.~\ref{fig:slp}, each transmitted symbol is associated with a Constructive Interference (CI) region. When the received symbol is within the CI region of the transmitted symbol, it will be accurately decoded as the transmitted symbol.
Let the $s^{\text{th}}$ transmitted symbol to user $k$ be $M_{\rm psk}$-modulated: $m_{ks}=m e^{j\phi_{ks}}$, the received symbol at the user can be equivalently expressed as 
\begin{align}
{y}_{ks}^{\mathrm{C}} &= \mathbf{ {h}}^{T}_{ks} \mathbf{x}_{s}m_{ks}+{n}_{ks}^{\mathrm{C}},
\label{yslpcbf} 
\end{align}
 where $\mathbf{ {h}}^{\Ttran}_{ks}= \mathbf{ {h}}^{\Ttran}_{k} e^{j(-\phi_{ks})}$.
To ensure ${y}_{ks}^{\mathrm{C}}$ falls within the CI region of $m_{ks}$, we must impose the following constraint on the transmitted symbol vector (see Fig. \ref{fig:slp} and \cite{8374931}):
\begin{align}
\left(\left|\text{Im}\left\lbrace\mathbf{ {h}}^{\Ttran}_{ks}\mathbf{x}_{s} \right\rbrace\right| - \left(\text{Re}\left\lbrace\mathbf{ {h}}^{\Ttran}_{ks}\mathbf{x}_{s} \right\rbrace\right)\tan(\psi) +\gamma^{'}{\sigma_{\mathrm{C}}}\frac{1}{\cos(\psi)}\right) \leq 0, \label{slpsc}
\end{align}
where $\gamma=\sqrt{\gamma}'$ with $\gamma$ being the minimum required communication SINR. Here, $\psi=\pi/M_{\rm psk}$ with $M_{\rm psk}$ being the order of the PSK modulation.
\begin{figure*}  
\begin{align}
\left[\mathbf{F}\right]_{\mathbf{l}\mathbf{p}} &= L \left( \dot{\mathbf{A}}_{\mathbf{l}}^{\mathrm{H}} \mathbf{C}^{-1} \dot{\mathbf{A}_{\mathbf{p}}} \right) \circ \left( \mathbf{B} \mathbf{V}^{\mathrm{T}} \mathbf{R}_{x} \mathbf{V}^{*} \mathbf{B}^{*} \right) +  L \left( {\mathbf{A}}^{\mathrm{H}} \mathbf{C}^{-1} \dot{\mathbf{A}_{\mathbf{p}}} \right) \circ \left( \mathbf{B} \mathbf{V}^{\mathrm{T}} \mathbf{R}_{x}\dot{\mathbf{V}_{\mathbf{l}}}^{*} \mathbf{B}^* \right)\nonumber\\
&+ L \left( {\dot{\mathbf{A}_{\mathbf{l}}}}^{\mathrm{H}} \mathbf{C}^{-1} {\mathbf{A}} \right) \circ \left( \mathbf{B} \dot{\mathbf{V}_{\mathbf{p}}}^{\mathrm{T}} \mathbf{R}_{x}{\mathbf{V}}^{*} \mathbf{B}^{*} \right) 
+ L \left( {{\mathbf{A}}}^{\mathrm{H}} \mathbf{C}^{-1} {\mathbf{A}} \right) \circ \left( \mathbf{B} \dot{\mathbf{V}_{\mathbf{p}}}^{\mathrm{T}} \mathbf{R}_{x}{\dot{\mathbf{V}_{\mathbf{l}}}}^{*} \mathbf{B}^{*} \right)\quad \forall \mathbf{l},\mathbf{p} \in \{\mathbf{\boldsymbol{\theta}^{\mathrm{t}}},\mathbf{d}^{\mathrm{t}}\}\label{first}\\
\left[\mathbf{F}\right]_{\mathbf{l}\mathbf{b}} &=L \left( \dot{\mathbf{A}}_{\mathbf{l}}^{\mathrm{H}} \mathbf{C}^{-1} {\mathbf{A}} \right) \circ \left( \mathbf{V}^{\mathrm{T}} \mathbf{R}_{x} \mathbf{V}^{*} \mathbf{B}^{*} \right)+ L \left( {\mathbf{A}}^{\mathrm{H}} \mathbf{C}^{-1} {\mathbf{A}} \right) \circ \left( \mathbf{V}^{\mathrm{T}} \mathbf{R}_{x} \dot{\mathbf{V}}_{\mathbf{l}}^{*} \mathbf{B}^{*} \right) \quad \mathbf{l}\in \{\boldsymbol{\theta}^{\mathrm{t}},\mathbf{d}^{\mathrm{t}}\}\\
\left[\mathbf{F}\right]_{\mathbf{b}\mathbf{b}} &= L \left( \mathbf{A}^{\mathrm{H}} \mathbf{C}^{-1} \mathbf{A}\right) \circ \left( \mathbf{V}^{\mathrm{H}} \mathbf{R}_{x}\mathbf{V}^{*} \right)\\
\mathbf{A} &= \left[\mathbf{a}_1 \quad \cdots \quad \mathbf{a}_L\right], \quad \mathbf{V} = \left[\mathbf{v}_1 \quad \cdots \quad \mathbf{v}_L\right], \quad \dot{\mathbf{V}}_{*} = \left[\dot{\mathbf{v}}_{1,*} \quad \cdots \quad \dot{\mathbf{v}}_{L,*}\right] \quad \dot{\mathbf{A}}_{*} = \left[\dot{\mathbf{a}}_{1,*} \quad \cdots \quad \dot{\mathbf{a}}_{L,*}\right],\\  
\mathbf{b} &= \left[b_1 \quad \cdots \quad b_L\right]^{\Ttran}, \quad \mathbf{B} = \text{diag}(\mathbf{b})\label{last}. 
\end{align}
\hrule
\end{figure*}

\section{Problem Formulation}
This section presents the proposed optimization framework and solution that designs the transmit symbol vector, balancing between sensing and communication performance:
\begin{subequations}
\begin{align}
\color{black}\text{(P1):~}&\color{black}\underset{\gamma^{'},t_{i},\{\mathbf{x}_{s}\}, \{\mathbf{R}_{\mathbf{X}_{s}}\}}{\text{maximize}}\,\,\,\,  \frac{-\rho\sum_{i=1}^{|\zeta|} t_{i}
}{\mathrm{NF}_{\mathrm{R}}} \,\,  +\frac{(1-\rho)\gamma^{'}}{\mathrm{NF}_{\mathrm{C}}},\nonumber\\
&\color{black}\begin{bmatrix}
\mathbf{F}&  \mathbf{e}_{i}\\  \mathbf{e}^{\Ttran}_{i} 
 &  {t_i}\\
\end{bmatrix} \succeq \mathbf{0},\label{c11}\\
&\begin{bmatrix}
 \mathbf{R}_{\mathbf{x}_{s}}& \mathbf{x}_{s}\\ 
 \mathbf{x}^{\Htran}_{s}& 1
\end{bmatrix}\succeq 0; \,\, \mathbf{R}_{\mathbf{x}_{s}} \,\, \succeq 0 \quad \forall s,\label{c1}\\
& \frac{1}{S}\text{tr}\left(\sum_{s=1}^{S}\mathbf{R}_{\mathbf{x}_{s}}\right)  \leq P_{\mathrm{t}},\label{c2}\\
&\eqref{slpsc}\label{c3}.
\end{align}
\end{subequations}
The objective function of $\text{(P1)}$ is a weighted combination of sensing and communication performance metrics with $\rho \in \left[0,1\right]$ being the weight factor. The scalars $\mathrm{NF}_{\mathrm{R}}$ and $\mathrm{NF}_{\mathrm{C}}$ are the normalization factors obtained by setting $\rho=1$ and $\rho=0$. \color{black}We use the trace of the CRB (which equals the inverse of the FIM) as the metric for sensing performance and the minimum SINR to evaluate communication performance. Here, $\{t_i\}$ represents the upper bound of the $i^{\text{th}}$ diagonal element of the CRB matrix through the Schur compliment $\mathbf{e}^{\Ttran}_{i}\mathbf{F}^{-1}\mathbf{e}_{i}\leq t_{i}$, represented as \eqref{c11}, where $\mathbf{e}_{i}$ is the $i^{\text{th}}$ column of the identity matrix $\mathbf{I}_{|\zeta|}$\color{black}. Equation \eqref{c1} represents the relation between $\mathbf{R}_{\mathbf{x}_{s}}$ and $\mathbf{x}_{s}$ through the relaxed constraint $\mathbf{R}_{\mathbf{x}_{s}} \succeq  \mathbf{x}_{s}\mathbf{x}^{H}_{s}$. The average transmit power is constrained to $P_{\mathrm{t}}$ using \eqref{c2}, whereas \eqref{c3} is the minimum SINR constraint. Note that $\mathbf{R}_{x}=\left({1}/{S}\right)\text{tr}\left(\sum_{s=1}^{S}\mathbf{R}_{\mathbf{x}_{s}}\right)$. \color{black}The objective function is convex since it is an affine function of the optimization variables $t_{i}$ and $\gamma^{'}$. The constraint \eqref{c11} is convex because $\mathbf{F}$ is an affine function of $\mathbf{R}_{x_{s}}$. Similarly, constraints \eqref{c1}-\eqref{c3} are affine functions involving $\mathbf{R}_{x_{s}}$ and $\mathbf{x}_{s}$, preserving convexity. Thus, (P1) is a convex optimization problem due to the affine nature of its objective function and constraints with respect to the variables $\mathbf{R}_{x_{s}}$,$\mathbf{x}_{s}$, $t_{i}$, and $\gamma^{'}$ and can be solved using general-purpose solvers, such as CVX.\color{black}
\vspace{-5mm}
\subsection{Baseline Problem Formulation with BLP}
Let $\mathbf{X}=\mathbf{W}\mathbf{D}$ where $\mathbf{W}=[\vect{w}_1,\vect{w}_2,\ldots, \vect{w}_K] \in \mathbb{C}^{N \times K}$ is the dual-functional  beamforming  matrix to be designed. Here, $\mathbf{D} \in \mathbb{C}^{K \times S}$ is the orthogonal data stream transmitted to the $K$ users: $(1/S)\mathbf{D}\mathbf{D}^{H}=\mathbf{I}_{K}$. The received symbol at user $k$ is represented as
\begin{equation} \label{eq:MU-MIMO-userk1}
    \hat{\mathbf{y}}_k = \vect{h}^{\Ttran}_{k}\mathbf{W}\mathbf{D}+ \vect{n}_k.
\end{equation}
Consequently, the received SINR at user $k$ is
written as
\begin{align}
  &  \text{tr}\left (  \mathbf{Q}_{k}  \mathbf{W}_{k} \right )-\gamma\left(\sum_{j\neq k}^{K} \text{tr}\left (  \mathbf{Q}_{k}  \mathbf{W}_{j} \right )\right) \geq \gamma \sigma^{2}_{C} \,\, \forall k,\label{sinrblp}
\end{align}
where $\mathbf{Q}_{k}=\mathbf{h}^{*}_{k}\mathbf{h}^{\Ttran}_{k}$, $\mathbf{W}_{k}= \mathbf{w}_{k}\mathbf{w}^{\Htran}_{k}$, and $\mathbf{W}_{j}= \mathbf{w}_{j}\mathbf{w}^{\Htran}_{j}$. Hence, the BLP equivalent of (P1) can be formulated as
 \begin{subequations}
\begin{align}
\color{black}\text{(P2):~} & \color{black}\underset{\gamma,t_{i},\{\mathbf{W}_{k}\}}{\text{maximize}}\,\,\,\,  \frac{-\rho\sum_{i=1}^{|\zeta|} t_{i}
}{\mathrm{NF}_{\mathrm{R}}} \,\,  +\frac{(1-\rho)\gamma}{\mathrm{NF}_{\mathrm{C}}},\nonumber\\
&\eqref{c11}, \eqref{sinrblp}\\
&\sum_{k=1}^{K}\mathrm{tr}\left(\mathbf{W}_{k}\right)  \leq P_{\mathrm{t}}. \label{p1.pt}
\end{align}
\end{subequations}
Note that the transmit correlation matrix, in this case, is computed as $\mathbf{R}_{x}=\sum_{k=1}^{K}\mathbf{W}_{k}$. Problem (P2) is non-convex due to multiplication between $\gamma$ and the interference term of \eqref{sinrblp}. \color{black}We apply the alternating optimization method described in \cite{babu2023multi} to solve (P2).\color{black}
\begin{figure}[ht!] 
\centering

\begin{subfigure}{\columnwidth} 
    \centering
    \includegraphics[width=0.7\linewidth]{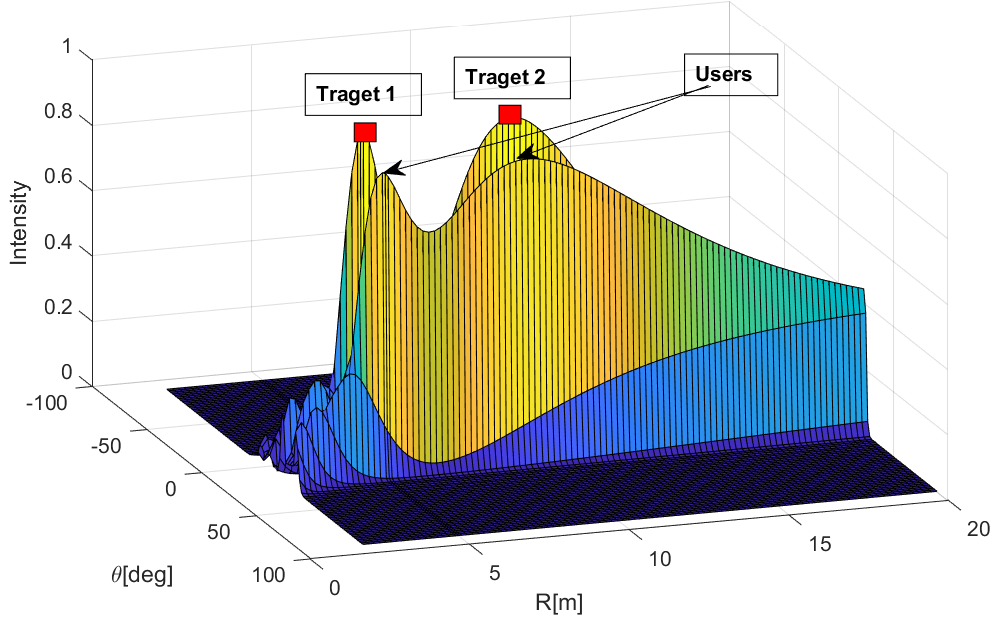} 
     \caption{$\theta^{\mathrm{t}}_{1}=\theta^{\mathrm{t}}_{2}=0^{o}$,$d^{\mathrm{t}}_{1}=5$m, $d^{\mathrm{t}}_{2}=10$m.}   
    \label{fig:sub1}
\end{subfigure}%
\hfill 

\begin{subfigure}{\columnwidth} 
    \centering
    \includegraphics[width=0.7\linewidth]{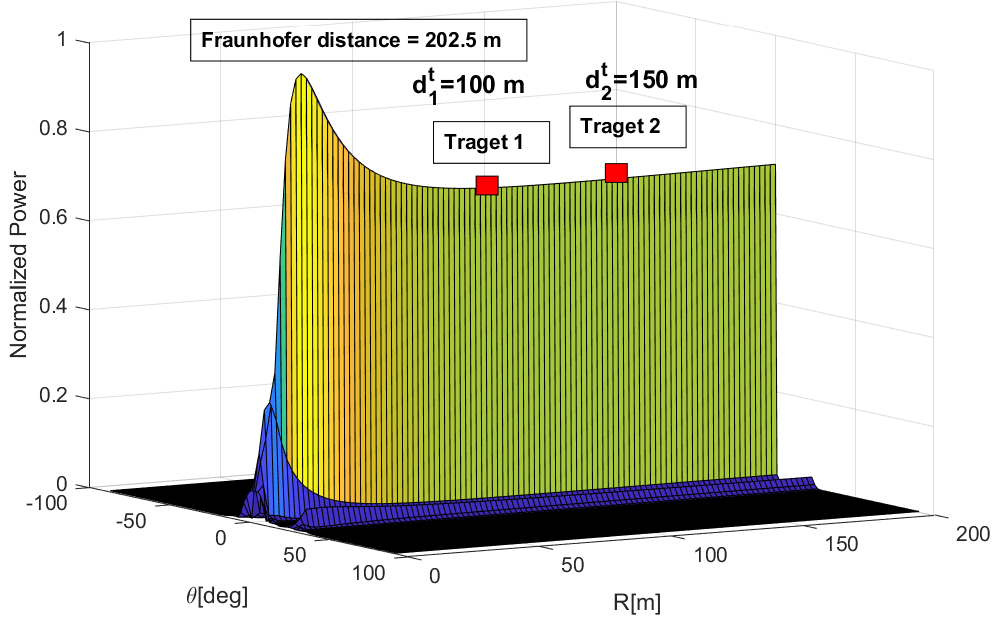} 
     \caption{\color{black}$d^{\mathrm{t}}_{1}=100\, \text{m}$,$d^{\mathrm{t}}_{2}=105$ m.}
    \label{fig:sub3}
\end{subfigure}
\caption{Beamfocusing when $N=201$, $f_{\mathrm{c}}=30$ GHz, $L=2$, $K=2$, $\psi=\pi/4$, $P_{\mathrm{t}}\,N/\sigma^{2}_{\mathrm{C}}=30$ dB. Beamfocusing disappears when either $\theta^{\mathrm{t}}\rightarrow 90^{o}$ or $d^{\mathrm{t}}\rightarrow d_{\mathrm{FA}}$.  }
\label{fig:result1}
    \vspace{-3mm}
\end{figure}
\begin{figure}
    \centering
    \includegraphics[width=0.75\columnwidth]{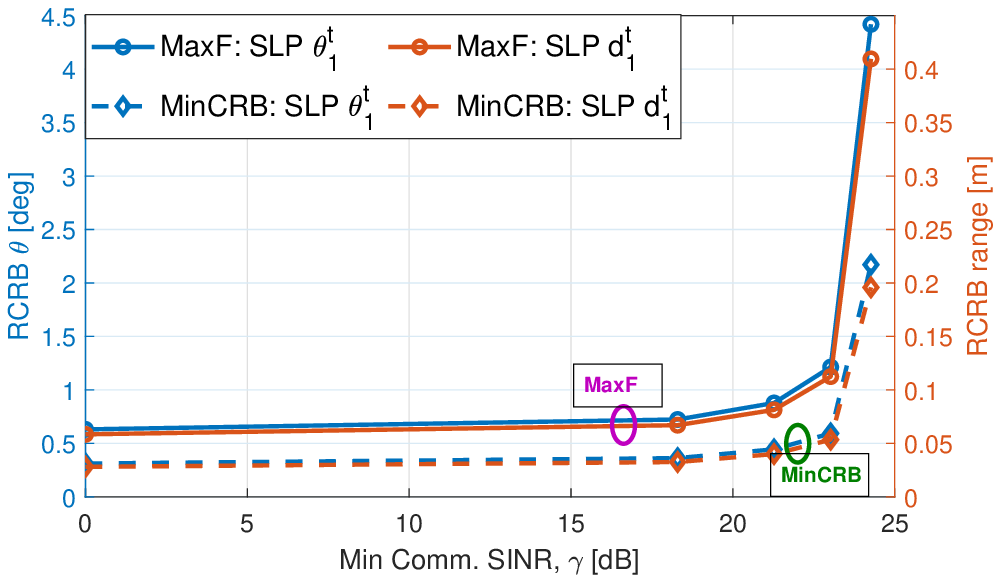}
    \caption{\color{black}RCRB Vs $\gamma$ comparison between maximizing the sum of the square root of the eigenvalues of $\mathbf{F}$ (MaxF) and minimizing the trace of CRB matrix (MinCRB), for $N=101$, $L=2$, $P_{\rm t}=10\, \rm\, dBm$ $\sigma^{2}_{\mathrm{C}}=0\, \rm dBm$, $K=2$, $\psi=\pi/4$. }
    \label{comparison}
\end{figure}
\begin{figure}
    \centering
    \includegraphics[width=0.75\columnwidth]{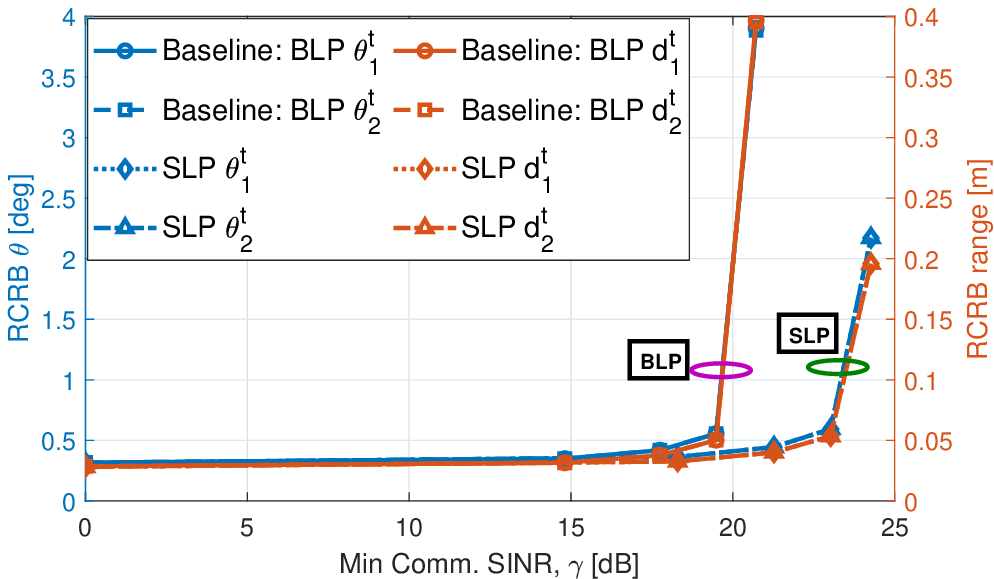}
    \caption{\color{black}Communication SINR vs.~Sensing RCRB trade-off for $N=101$, $L=2$, $P_{\rm t}=10\, \rm\, dBm$ $\sigma^{2}_{\mathrm{C}}=0\, \rm dBm$, $K=2$, $\psi=\pi/4$. }
    \label{fig:result2}
    \vspace{-6mm}
\end{figure}

\vspace{-8mm}
\subsection{Multiple Signal Classification (MUSIC) Algortihm}
We use MUSIC algorithm to estimate the angle and distance using the received echo samples obtained after transmitting $\mathbf{X}$ obtained from (P1). The MUSIC algorithm works by exploiting the structure of the eigenvectors of the sample covariance matrix:
\begin{align}
\widehat{\vect{R}}_L = \frac{1}{S} \sum_{s=1}^S \mathbf{y}^{\mathrm{R}}_s(\mathbf{y}^{\mathrm{R}}_s)^{\Htran}.\label{Rl}
\end{align}
Given the number of targets $L$, we first construct  the noise-subspace matrix $\widehat{\vect{U}}_{\rm n}\in \mathbb{C}^{N \times (N-L)}$ whose columns are the eigenvectors of $\widehat{\vect{R}}_L$ corresponding to the smallest $(N-L)$ eigenvalues. The $2$D MUSIC spectrum is generated as \cite{Stoica2005_book}
\begin{align} \label{eq:MUSIC-spectrum}
S(d,\theta)=\frac{1}{\vect{a}^{\Htran}(d,\theta)\widehat{\vect{U}}_{\rm n}\widehat{\vect{U}}_{\rm n}^{\Htran}\vect{a}(d,\theta)}.
\end{align}
The targets' locations are ascertained by identifying $L$ peaks within the $2$D MUSIC spectrum.
\section{Numerical Evaluation}
 This section explains the simulation outcomes. 
 We consider a linear array with a half-wavelength spacing, operating at $f_{\rm c}= 30$ GHz. This antenna arrangement gives a Fraunhofer distance of $80$ m. We analyze a scenario where there are $K = 2$ users, and $ L = 2$ targets. We consider the challenging scenario where both targets are positioned in the same angular direction but at varying distances, specified as $[5\, \text{m}, 0^{\circ}]$ and $[10\,\text{m}, 0^{\circ}]$. The users are located at $[10\,\text{m}, 22.5^{\circ}]$ and $[15\,\text{m}, 22.5^{\circ}]$. Note that this arrangement places all users and targets in the near-field of the transmit and receive antenna arrays. We define the array SNR (ASNR) as $P_{\mathrm{t}}N^{2}/\sigma^{2}_{\mathrm{R}}$ and use $\rm ASNR = 10$ dB.    

Fig. \ref{fig:result1} presents the beampattern we obtained when $\gamma^{'}=1$. It shows that the obtained solution focuses the maximum intensity toward the exact locations of the targets. Furthermore, it is observed that as the communication demand increases, the focus of the beam progressively shifts in the direction of the users. Another important finding is that when the target's azimuth angle aligns more towards the end-fire direction, we observe a considerable degradation in the beam-focusing capability. 
As the signal's source is closer to the end-fire direction, the focusing capability diminishes \cite{2024_Kosasih_TWC}.
The decrease in focusing capability is also evident when the targets are closer to the Fraunhofer array distance ($d_{\rm FA}=(N\cdot \Delta)^{2}/\lambda$). This corresponds with the findings in \cite{2021_Björnson_Asilomar}, where the finite beam depth disappears when the propagation distance exceeds ten percent of Fraunhofer array distance, significantly smaller than $d_{\rm FA}$. Hence, even though the targets remain in the near-field, the focusing capability diminishes as it surpasses $d_{\rm FA}/10$. 

Additionally, the MUSIC algorithm utilizes $\hat{\mathbf{R}}_{L}$, calculated as per equation \eqref{Rl}, to select the pairs of angle ($\theta$) and distance ($d$) from grids sampled with a resolution of $2000 \times 2000$ points, that maximize the MUSIC spectrum value given in \eqref{eq:MUSIC-spectrum}. The estimated values are $\hat{\theta}^{\mathrm{t}}_{1}=0.5239\cdot 10^{-3}$, $\hat{\theta}^{\mathrm{t}}_{2}=-0.5239\cdot 10^{-3}$, $\hat{d}^{\mathrm{t}}_{1}=4.93$ m, and $\hat{d}^{\mathrm{t}}_{1}=10.31$ m. 

\color{black}Fig. \ref{comparison} shows that the considered performance metric outperforms other sensing-related objective functions, such as maximizing the sum of the square root of the eigenvalues of the FIM. As shown in the figure, minimizing the trace of CRB matrix yields a lower estimation error compared to maximizing the sum of eigenvalues of the FIM. The plots are obtained by varying $\rho \in \left[0, 1\right]$\color{black}.

Fig. \ref{fig:result2} compares the SLP and BLP performances. It shows the minimum communication SINR versus the root CRB (RCRB) to estimate the target angle (indicated by blue curves) and range (indicated by red curves). At lower communication SINR requirements, SLP and BLP yield similar outcomes. This similarity arises due to the minimal co-channel interference experienced at low SINR levels, which means there is less interference for SLP to utilize constructively to improve SINR. As the communication SINR demand increases, BLP becomes infeasible after a certain threshold due to excessive signal leakage between user locations. In contrast, SLP can operate at higher SINR values by leveraging the signal leakage. Nonetheless, at very high communication demands, the power initially allocated for sensing must be redirected to meet these communication demands, leading to a rise in the RCRB value.

\section{Conclusion}
We developed a framework to optimize the transmit symbol vector for simultaneously sensing multiple targets and serving multiple users in near-field ISAC, leveraging SLP. The goal was to balance reducing target estimation errors and enhancing the minimum achievable SINR for communication while adhering to a maximum total power constraint. The problem we addressed was convex, allowing us to derive a solution that proved effective when applied within the 2D MUSIC algorithm for estimating the positions of the targets. Our findings show that SLP offers superior performance compared to BLP. Additionally, SLP can function effectively in higher SINR environments than BLP. It was also noted that the focusing ability of the beam diminishes when targets or users are positioned close to the array's end-fire direction or near the Fraunhofer distance. \color{black}We designate uniform planar array (UPA) studies in multi-cell scenarios
as the focus of our future work. \color{black}.

\bibliographystyle{IEEEtran}

\bibliography{IEEEabrv,myBib}

\end{document}